# Who Acts, Who Knows, Who Answers? A Corpus-Assisted Discourse Analysis of Agency, Epistemic Responsibility, and Accountability in Generative AI Higher Education Research

Biranchi Poudyal

*Faculty of Arts and Society, Charles Darwin University, Australia*

ORCID: https://orcid.org/0000-0002-7210-5480

## Abstract

Generative artificial intelligence (GenAI) is increasingly described in higher education as a tool, collaborator, evaluator, proxy, and infrastructure. These labels are not neutral: they shape who is seen as acting, knowing, and, crucially, being answerable when someone uses AI. This study examines how scholars distribute such responsibility across 366 English-language titles and abstracts published between November 2022 and 14 August 2026. A corpus-assisted discourse analysis combined frequency counts, five-word collocations, actor–predicate associations, obligation-clause coding, and heuristic analysis of passives, nominalisations, and reporting metonymies. The final corpus contained 91,405 tokens, drawn from 8,734 unique records screened after deduplication. AI was the most frequently named actor, appearing 2,050 times and entering 447 predicate associations, most involving action. Students, by contrast, were more often associated with knowing, judging, and verifying, yet no sentence ever made AI explicitly responsible. Responsibility fell largely on educators, institutions, and policy, or disappeared through passive and nominalised constructions. Even frequent phrases such as "responsible AI" and "responsible use" rarely identify who is accountable for what. The findings reveal a persistent gap between giving AI functional agency and assigning normative accountability.

## 1. Introduction

Generative artificial intelligence is entering higher education through an ambiguous vocabulary. An LLM is a tool when it edits your writing, a partner when it continues your conversation, an agent when it executes the next operation, a teacher when it gives you explanations, and a substitute when it completes tasks that are assessed as belonging to students. The terms used above are not synonyms. They produce different presuppositions about whose action is generated, whose judgment is exercised, whose knowledge is sufficient to support a claim, and whose fault lies if the result of using the LLM misleads someone, or if an assessment indicates that an intended ability has been lost.

Most early discussions of GenAI in higher education centered on academic integrity, detection, and appropriate use. Perkins (2023) argued that current integrity categories for LLMs needed revision. Similarly, Crawford et al. (2023) distinguished between useful AI help and authorship. Both papers established ethical standards for the use of AI by human researchers. Sullivan et al. (2023) identified significant educational potential in the use of ChatGPT while also flagging the many risks to learning and academic integrity. Research since then has grown considerably more differentiated. Student research now examines epistemological beliefs, critical thinking skills, self-disclosure, cognitive offloading, and decision-making around revisions (e.g., Jiang et al., 2026; Liu et al., 2026; Qu & Wang, 2026; Urban et al., 2025). Policy research, meanwhile, has moved from prohibitions toward guidelines, regulations, and more equitable methods of assessing evidence (Alotaibi et al., 2026; Sporrong et al., 2025; Ucan, 2026).

This growth pushes research beyond a single question: Is GenAI used? - toward a harder one: what human capability remains observable once GenAI is used? Several strands feed this inquiry. Measurement research maps dimensions of learner agency and distributed critical thinking (Cong-Lem, 2025; Dai et al., 2026); comparative research documents differences between perceived human and AI agency across educational settings (Essien et al., 2026); writing research reports both empowerment and the risk that system assistance replaces epistemological labor (Moorhouse et al., 2025); and cognitive-offloading research distinguishes productive delegation from dependency (Si et al., 2026). At the institutional level, policy analyses show variability in institutional responses to academic integrity and governance (De Maio, 2024), and conceptual work suggests that large language models can reconfigure agency through infrastructural arrangements as well as direct user experience (Lindebaum et al., 2026). Pedagogical proposals accordingly emphasize citation literacy, self-disclosure, and the capacity to justify AI-assisted work (Karam, 2026; Madsen & Silva, 2026). While these areas of research connect through several themes, the terminology used to describe agency and responsibility rarely receives analysis as a recurring theme in the discourse.

The language of this scholarship is itself part of the problem. A statement like “ChatGPT evaluates a response” semantically activates a system. The phrase “Students must verify the output” obliges learners to act. “Responsible AI should be implemented” uses a normative adjective, a passive-voice construction, and names no actor to carry out implementation. Each phrase individually may make sense in context, but together they build norms that separate systems - the doers of creation, production, and transformation - from people, who bear verification and consequence. At the same time, “active AI” language can offer convenient shorthand, and dialogic interaction with systems can encourage rather than inhibit human judgment. The analytical task here, therefore, is not to strip anthropomorphic language from scholarly description, but to examine where that language appears and how it relates to explicit accountability.

Reviews of prior GenAI research identify applications, attitudes, and benefit-risk tradeoffs, but few treat the grammar of the literature as evidence in its own right. Perkins and Roe (2024) demonstrated the value of analyzing repeated lexical patterns across policy recommendations to trace technological threats to academic-integrity policy, using a corpus-based methodology. Unlike reviews that focus on applications, attitudes, or risk, this review specifically focuses on GenAI scholarship in higher education. More precisely, it analyses how GenAI scholars allocate action, epistemic work, and responsibility in titles and abstracts

- two locations that remain among the most consistently visible and available representations of an author's argument.

This scope of inquiry occupies a distinct position within larger bodies of research on learner-AI interaction, assessment task design, and institutional accountability. Interactional accounts describe how responsibility for learner-AI use evolves over the course of a learner's engagement; task-design accounts distinguish the limited assistance that those systems provide from the proxy performance those same systems achieve. This review does neither. Rather than proposing new frameworks or additional levels of abstraction, it addresses the issue at the level of the discipline, examining how GenAI scholars formulate responsibility relations and whether their recurring formulations preserve the distinction between operant performance and accountable judgment.

The study addresses four questions:

1. How does peer-reviewed GenAI education scholarship linguistically represent learners, educators, AI systems, and institutions as actors, knowers, and evaluators?
2. Which grammatical, lexical, and conceptual patterns attribute, distribute, obscure, or displace epistemic responsibility?
3. How do these patterns vary across empirical, conceptual, review, and policy-oriented publications, and across tool, interlocutor, proxy, and infrastructural representations of GenAI?
4. To what extent does the literature preserve or blur the asymmetry between AI's functional agency and human or institutional accountability?

## 2. Agency, Epistemic Work, and Accountability as Discourse

### 2.1 From Grammatical Activation to Answerability

The term "agency" should not be reduced to a single conceptualization; it operates in at least three senses. Grammar-based agency refers to how a clause depicts participants, who function as actors, sensors, speakers, or decision-makers. Function-based agency concerns a system's ability to perform activities that alter other activities. Epistemically based agency refers to an individual's involvement in the creation, evaluation, and justification of knowledge. A language model can produce function-based effects without thereby possessing lived experience, ethical accountability, or the capacity to accept consequences. This differentiation between types of agency resembles Floridi and Sanders's (2004) distinction between artificial agency and human moral accountability, as well as Coeckelbergh's (2020) account of relational responsibility attribution. It also cautions against assuming that a grammatically indicated subject represents an ontologically determined property.

Epistemic work involves defining problems, developing alternative solutions, interpreting evidence, comparing accounts, validating references, modifying claims, and determining what will be accepted. Multiple human and computational contributors may take part in this work, but contributing to it does not confer equal authority. A model can produce a reasonable explanation; a learner can accept, dispute, or modify that explanation; an instructor can set the standards for evaluation; an institution can ensure that a final assessment demonstrates a student's possession of a specific graduate capability. Although the contributors depend on one another, they do not hold the same level of authority.

Epistemic responsibility involves answerability as well as participation: it addresses who is obligated to justify the creation, acceptance, and use of a particular claim. A genuine responsibility relationship involves more than a single adjective - it has a bearer, an epistemic object or decision, an anticipated action, a standard, an audience owed justification, and some form of consequence or redress if the responsibility goes unmet. "Responsible use" names a desired quality, but it does not necessarily name the relationship described above. "The outputs should be verified" names an action. However, it may not identify who acts, against what criteria the output will be verified, or what consequence follows a failure to verify.

### 2.2 Transitivity, Modality, and Social Actors

Systemic Functional Linguistics views clauses as representations of processes and participants (Halliday & Matthiessen, 2014). Material processes show participants acting; mental processes illustrate sense-making, knowing, and judging; verbal processes establish speakers and sources. Transitivity analysis does not determine whether a participant holds philosophical agency; rather, it illustrates the recurring grammatical options through which agency becomes possible.

Social Actor Analysis adds further considerations: activation/passivation, inclusion/exclusion, and impersonalization (van Leeuwen, 2008). When we say "the AI generates feedback," we activate a system. When we say "feedback is generated," we remove the agent. When we say "policy requires disclosure," we metaphorically activate an organizational artifact, collapsing all the administrative personnel and approval mechanisms behind a policy into a single abstract noun. Nominalizations like governance, verification, and accountability turn processes into transferable objects. Nominalizations can help build new concepts, but they can also obscure who is governing, verifying, or answering.

Modality matters here too: modal verbs such as must, should, need to, required to, responsible for, and accountable for express obligation. That obligation may be stated explicitly ("Educators should redesign assessment"), collectively ("Stakeholders must collaborate"), metaphorically ("Curriculum should develop judgement"), or not at all ("Guidance is required"). Corpus-assisted discourse analysis pairs corpus-based frequency identification with the interpretive focus central to critical discourse analysis (Baker et al., 2008; Fairclough, 1992). Frequency identification establishes patterns but not motivations, and concordance reading resists assigning a repeated word a single meaning without reading it in context.

### 2.3 Tool, Interlocutor, Proxy, and Infrastructure

Four operational frameworks guide comparison within this project. Tool discourse presents GenAI as limited assistance or an object of use; it typically emphasizes adoption, integration, and human control. Interlocutor discourse positions GenAI as a conversational partner, tutor, collaborator, or co-developer, supporting educational purposes such as longitudinal writing and design projects, in which students receive, revise, or reject AI contributions (Willems et al., 2025; Tan, 2026). Proxy discourse concerns substitution: GenAI completes intellectual tasks credited to, or expected of, learners. Proxying is not intrinsically unethical - students can engage in critiquing or evaluating AI-produced texts - and it is not harmful if students remain capable of demonstrating the abilities that assessments based on their own work are meant to show. Work on ghost students and assessment presence addresses this issue clearly (Bozkurt et al., 2026).

Infrastructure discourse positions GenAI within policy, assessment, research, data, and institutional governance. It shifts focus away from the original input prompt toward the systemic arrangements that render certain behaviors acceptable, recognizable, and significant (Alotaibi et al., 2026; Wu, 2026a/b).

These four categories are descriptive rather than prescriptive - interlocutor framing can mask the delegation of judgment while tool framing supports it; infrastructure framing can just as easily assign responsibility clearly or obscure it behind policy terminology - but the categories remain useful for comparative analysis. They let researchers assess whether the same description of GenAI shifts which verb forms attach to action, responsibility, and obligation.

## 3. Method

### 3.1 Design and Scope

This study used corpus-assisted discourse analysis of peer-reviewed titles and abstracts, along with transparent evidence identification and screening. It drew on public documentary materials and bibliographic metadata. No participants were recruited; therefore, no private participant-level data were collected. Titles and abstracts may report participant research, but they serve here as the public-facing units of analysis.

Eligibility criteria were defined as follows: the eligibility period ran from 1 November 2022 to 14 August 2026. Eligible records were limited to English-language journal articles, reviews, and peer-reviewed conference papers focused on GenAI in higher education. To qualify, each record had to contain at least three of eight domains: action, knowledge, judgment, verification, responsibility, attribution, governance, and agency. Applications outside education, school-only research, technical models of GenAI without educational discourse, unreviewed commentary, GenAI references in non-analytical contexts, and records lacking analytically comparable or retrievable text were all ineligible.

Titles and abstracts served as the common textual unit largely because of legal constraints on retrieving full-text materials across databases, which made full texts unavailable consistently. Combining full texts with abstracts elsewhere would have confused genre, length, and patterns of discussion. Because of these limitations, the study makes claims about the public argumentation found in titles and abstracts, not about the entire article.

### 3.2 Search and Duplicates

Searches were conducted independently across four sources: Scopus, Web of Science Core Collection, IEEE Xplore, and SpringerLink. The master search logic combined three blocks: GenAI terms, higher-education terms, and agency/responsibility terms. The third block included agency, autonomy, authorship, accountability, responsibility, epistemic trust/authority/trustworthiness, academic integrity, oversight/human oversight/evaluative judgment, and verification/governance/decision-making. Database syntax, fields, filters, date ranges, and count totals appear in the supplementary methods.

Based on title, abstract, and keyword fields, Scopus yielded 3,590 records. From topic records alone, Web of Science yielded 2,377 records. IEEE's field nesting and term limits required a documented translation of metadata scope; because Scopus and Web of Science already covered conference literature, the protocol limited the IEEE export to 1,300 journal records. SpringerLink's full-text search returned 16,271 records. Because export limits capped downloads at 1,000 records per session, the complete 2022–2023 results plus the relevance-ranked top 1,000 for each year of 2024–2026 generated 3,908 Springer records - a substantial coverage restriction that makes the Springer results a non-random, non-exhaustive sample.

Twelve export files contained 11,175 rows in total. Rows were standardized, and duplicates were removed based on an exact normalized DOI or, where no DOI existed, an exact normalized title. Where a title lacked a DOI but matched a title already carrying one, it was added to that DOI group; distinct DOI groups were never merged simply because titles matched. Three malformed Web of Science/IEEE DOIs were repaired by cross-referencing export information. Removing duplicates eliminated 2,441 rows, leaving 8,734 unique records.

**3.3 Screening and Corpus Freeze**

Before interpretive review, a first-pass transparency check determined whether a record showed focus on GenAI terminology, an educational context, at least three of the eight discourse domains (action, knowledge, judgment, verification, responsibility, attribution, governance, agency), and at least two abstract sentences showing both an actor and a domain reference. A centrality score combined title-weighted focus expressions tied to GenAI, capped abstract hits (records with at least two actor/domain-referencing sentences), domain breadth (the number of domain expressions in a record), and actor-domain sentence count. Records scoring twenty or above proceeded to interpretive review; a sensitivity path also admitted records whose titles clearly combined GenAI, higher education, and agency/responsibility, regardless of centrality score.

Reviewing these records excluded sixteen as clear false positives - wireless networking, antimicrobial-governance advertising, and K-12 case studies where GenAI was secondary to another issue - and forty-eight Springer-only titles that lacked a corresponding abstract. The final corpus consisted of 366 documents: eight from 2023, thirty-five from 2024, ninety-seven from 2025, and two hundred twenty-six from 2026. The PRISMA-informed flow (Page et al., 2021) traces this path: 11,175 exported rows; 2,441 duplicate rows removed; 8,734 unique rows entered screening; 8,368 records excluded at screening; and 366 final inclusions.

**3.4 Corpus Preparation and Analysis**

All text was normalized and tokenized using a Unicode-aware expression. One abstract contained an intact English abstract immediately followed by 607 literal "(sic)" markers, an artifact of corrupted non-Latin export suffixes; the original record was preserved, and only the corrupted suffix was removed from the analysis copy. The clean corpus contained 91,405 tokens, 7,651 types, and 3,981 sentences.

Five actor dictionaries denoted AI, students, educators, institutions, and providers/developers. Eight strict verbal predicate dictionaries denoted action, knowledge, judgment, verification, responsibility, attribution, governance, and agency. An association between an actor and a predicate was recorded when an actor preceded a verbal predicate in the same sentence, within six intervening words. Nominal proximity was intentionally excluded: "AI governance" was not recorded as evidence that AI governs. The method is a rule-based approximation rather than a dependency parse.

Collocations were computed over symmetric five-word windows and reported only when they occurred at least 5 times across three documents, alongside raw frequency, document dispersion, and Log Dice. Because no external corpus adequately matched the focal corpus in period, discipline, and genre, abstract keyness was not calculated.

Obligation expressions - must, should, need to, ought to, require, responsible for, accountable for - were extracted and coded by concordance for their bearer. Separate grammar counts flagged candidate passive constructions (be plus past participle), by-agent phrases, responsibility-related nominalizations, and reporting metonymies. These grammar counts are diagnostic indicators, not perfect annotations.

Publication type was classified as empirical, conceptual/position, review/synthesis, or policy/governance. The author then reviewed the classifications and output the entire register; decisions, dictionaries, and scripts accompany the article.

### 3.5 Rigor, Reflexivity, and Research Integrity

The study's focus on responsibility risks interpreting every AI-active construction as displacement. Three safeguards addressed this risk. First, grammatical activation was never treated as consciousness or moral agency. Second, automated verification and productive human-AI collaboration were retained as negative cases rather than discarded. Third, the operational categories were treated as non-hierarchical, and explicit actor predicates were analysed separately from nominal proximity. Following the corpus freeze, the author manually rechecked all counts, preserved source-row provenance for every row, and verified individual DOI metadata against Crossref records. However, follow-up verification was not always possible.

## 4. Results

### 4.1 Corpus Composition and Lexical Priorities

The corpus contained 239 empirical articles (65.3%), 47 policy/governance analyses (12.8%), 42 reviews (11.5%), and 38 conceptual/position papers (10.4%). "Tool" was the main concept in 146 records (39.9%), followed by "infrastructure" in 126 (34.4%), "interlocutor" in 48 (13.1%), and "proxy" in 46 (12.6%). The corpus grew rapidly toward 2026, with a partial year alone accounting for 61.7% of records.

Words such as "academic", "learning", "integrity", "ethical", "critical", "thinking", "responsibility", "assessments", "governance", "agency", and "epistemology" were prominent, occurring 758, 539, 461, 431, 378, 267, 255, 229, 227, 222, and 151 times respectively. Forms of verification appeared 158 times, responsibility/responsibilities 104 times, authorship 96 times, and accountable/accountability 72 times.

A significant lexical asymmetry appears in repeated phrases. "Responsible AI/GenAI" appears 83 times across 59 documents, and "responsible use" occurs 66 times across 44 documents. "AI responsibility" never appears as an exact phrase, and neither does "student responsibility". "Institutional responsibility" appears twice in one record, and "human responsibility" appears once. "Educator responsibility" appears four times, all in a single policy record. Only one sentence explicitly links AI and accountability - stating that AI is not accountable. So, although the literature often invokes responsibility, it typically names the technology or practice responsible rather than describing an accountable relationship.

**4.2 Who Acts, Knows, and Verifies?**

AI expressions were the most frequent actor labels (2,050), followed by students (1,385), institutions (545), educators (389), and providers/developers (25). AI also received the largest number of strict actor–predicate associations (447), but their semantic distribution was narrow.

**Table 1. Strict actor–predicate associations**

| Actor | Act. | Know. | Judg. | Verif. | Resp. | Attr. | Gov. | Agency | Total |
|---|---|---|---|---|---|---|---|---|---|
| AI | 352 | 5 | 13 | 11 | 0 | 31 | 17 | 18 | **447** |
| Student | 222 | 26 | 35 | 23 | 3 | 22 | 7 | 16 | **354** |
| Educator | 46 | 6 | 11 | 2 | 0 | 3 | 10 | 3 | **81** |
| Institution | 55 | 2 | 9 | 3 | 1 | 7 | 12 | 2 | **91** |
| Provider/developer | 4 | 0 | 0 | 1 | 0 | 0 | 1 | 0 | **6** |

*Note. Act. = action; Know. = knowledge; Judg. = judgement; Verif. = verification; Resp. = responsibility; Attr. = attribution; Gov. = governance.*

Action made up 78.7% of all references to AI - a system that creates, supports, changes, produces, or assists. Knowledge, judgment, and verification together made up only 29 AI association references. Students, by contrast, received predicates for the same three epistemic domains on 84 occasions, positioning them as able to understand, judge, interpret, check, and verify their own work, and to accept, revise, or reject AI output. Jiang et al. (2026), for example, distinguished several levels of learner choice regarding revisions based on degrees of learner control, while Urban et al. (2025) linked the degree of ChatGPT-content integration into educational materials to both epistemic beliefs and metacognitive accuracy.

AI verification predicates were not absent: eleven references across ten separate documents identified a system or layer of automation performing checks, reviews, or validation functions. These are important counterexamples, since they show that students do not completely lose decision-making capacity, but none of the constructions held the validating system accountable for potential consequences. Instead, each construction presented validation as just another action for students to evaluate.

The scarcity of providers was itself a finding. Commercial developers/vendors appeared only 25 times and entered six strict predicate associations. As a result, the organizational actions that shape data, model behavior, and platform accessibility appear far less often to be organizations' own actions than to be characteristics of "AI". Institutions appeared more often and specialised comparatively in governance, but their 91 strict predicates remained far below the system and learner totals.

**4.3 Collocation and Operational Representations**

Among AI's strongest dispersed collocations were tools (247 window occurrences across 130 records), responsible (163 across 99 records), literacy (153 across 73 records), academic (191 across 119 records), integration (135 across 86 records), learning (136 across 84 records), ethical (128 across 75 records), integrity (107 across 69 records), and governance (89 across 38 records). These collocations position systems simultaneously within an instrumental vocabulary and within programs that frame the norms around them.

Collocations for students positioned them more prominently within an agential vocabulary: agency (101 across 41 records), learning (108 across 65 records), academic (109 across 67 records), thinking (70 across 42 records), critical (72 across 49 records), engagement (48 across 34 records), integrity (54 across 40 records), and writing (44 across 22 records). The difference is not simply that AI acts while students do not - students do act. However, their activity more often sits within development, judgment, ethics, conduct assessment, and the maintenance of an accountable identity.

Operational representation shifted the distribution significantly. Tool records averaged 119.2 AI-action predicates per hundred records, proxy records averaged 89.1, interlocutor records averaged 87.5, and infrastructure records averaged 75.4. Tool-based discourse, then, does not necessarily minimize the extent to which systems act; it often simply represents what tools enable, create, or change.

Verification-judgement combinations were greatest in proxy discourse (26.1 associations per hundred records), consistent with the need to determine whether a submitted performance remains the learner's own. Tool discourse averaged the second-highest rate at 20.5 associations per hundred records, followed by interlocutor discourse at 16.7. Infrastructure discourse trailed well behind, averaging only 6.3 associations per hundred records.

Infrastructure discourse also averaged the highest rate of responsibility-nominalizations, at approximately 3.98 per document; proxy discourse followed at approximately 3.59; interlocutor discourse at approximately 2.73; and tool discourse at approximately 2.42. Governance, integrity, accountability, and oversight cluster most densely wherever institutional arrangements are the primary focus, though abstraction did not always assign responsibility. Policy, in other words, can appear to act while the administrators who enact and enforce it stay invisible.

Important counterevidence appears in interlocutor texts. Tan (2026) found that students did not perceive AI writing as a loss of personal agency; Willems et al. (2025) found that students used tool-teammate-deliberate non-use strategies; Rauschenbach (2026) found that AI served as a teaching partner within a design that preserved critical evaluation. These examples show that dialogue or collaborative wording does not automatically replace evaluation; what matters is whether students still possess and are required to use justification for their selections, revisions, or rejections.

### 4.4 Who Answers?

The corpus contained 166 obligation or answerability expressions: 37 named institutions or policy actors, 20 named educators, and 30 used assessment, curriculum, policy, or pedagogy metonymically in ways coded as institution/educator. Researchers or authors received 18 obligations, and students received six. Thirty-three clauses were impersonal or left the bearer unspecified.

**Table 2. Coded bearer of obligation expressions**

| Bearer | Clauses |
|---|---|

| Institution/policy | 37 |
|---|---|
| Institution/educator metonymy | 30 |
| Educator | 20 |
| Researcher/author | 18 |
| Student/learner | 6 |
| AI/system role constraint | 12 |
| Provider/developer | 2 |
| Other/shared human actors | 3 |
| Research caveat or non-prescriptive mention | 5 |
| Impersonal/unspecified | 33 |
| **Total** | **166** |

Twelve distinct AI/system clauses explained how systems ought to be used, framed, or restricted. None assigned moral or institutional consequences to systems. Crawford et al.'s (2023) early authorship policy makes this distinction clear: AI can assist authors but cannot be considered an accountable author. Recent "Responsible AI" literature likewise differentiates among the responsibilities of developers, deployers, governments, and users rather than assigning accountability within the model itself (Atamhenwan et al., 2026).

Institutional responsibilities included updating policies, redesigning assessments, providing guidance, embedding AI literacy training, and establishing procedural frameworks for disputing decisions regarding student performance. Sporrong et al. (2025) show why clear assignment matters: policy can assign educators responsibility for implementation without giving them equivalent control over institutional conditions. Ucan (2026) similarly argued that outputs from AI detection tools should never independently warrant punitive measures. Wu's (2026a, 2026b) research on governance relocates student accountability from individual learners alone to courses and doctoral education programs.

Student obligations were relatively few, given the size of the integrity vocabulary, largely because many were written indirectly: verification, critical thinking, and disclosure appeared as desired competencies rather than direct sentences naming responsible students. In empirical studies, though, those competencies remained tied to learner performance. Qutieshat et al. (2026) reported inconsistent verification/disclosure practices among dental schools; Jiang et al. (2026) connected agency to revision decisions; Liu et al. (2026) connected cognitive offloading and epistemic trust transfer to lower learning accountability. Responsibility was thus often individualized through measurement, regardless of whether it was named explicitly as "student responsibility".

**4.5 Passives, Nominalisation, and Metonymic Disappearance**

The heuristic rules identified 800 passive constructions, 780 without an agent reference, 1,152 responsibility-related nominalizations, and 241 reporting metonyms. These counts do not offer a precise measure of grammar. However, the scale still draws attention to repeated linguistic formulations - for example, guidance that calls for assessments to be redesigned and outputs to be verified without ever naming who the responsible actor is.

Nominalization proved more significant than simply deleting actors from sentences. Nominalizations of governance, responsibility, verification, authorship, and accountability enable researchers to build a portable set of norms within their field. At the same time, naming these concepts creates an impression that responsibility exists, simply because they are named. "Governance is needed", for example, does not say who convenes meetings, who provides funding, or who enforces governance. "Verification is essential" does not say who verifies claims against sources. "Transparency" does not specify who reveals what information to whom.

Reporting metonyms are a fairly frequent academic convention (the study demonstrates, findings indicate) and do not always signal problematic evasion. Their significance lies in what they reveal through comparison: how much agency accrues to the researcher's own text or to AI systems, versus how little direct identification is attributed to the organizations or roles that bear the consequences. Institutions do not disappear entirely here; institutional names appear often in "infrastructure" texts. What remains lacking is clarity about each institution's actual role in decision-making.

## 5. Discussion

### 5.1 Functional Agency Without Normative Equivalence

The corpus shows a significant and largely implicit asymmetry. AI is grammatically active - it generates, supports, produces, transforms, and at times evaluates or verifies - a clear functional agency. However, the system is never portrayed as an epistemically authoritative entity whose claims warrant themselves, or as an accountable agent capable of accepting sanctions, repairing institutional damage, or justifying an educational decision. No "strictly responsible" predicate has ever been assigned to AI, and its one direct accountability clause explicitly states that no accountability applies.

This finding avoids two extreme readings. It does not portray GenAI as merely a passive instrument, since the materiality of GenAI operations modifies writing, provides feedback, assesses students, and alters access to information. Nor does it equate grammatical activation with personhood. The literature typically employs practical agency language while still retaining human answerability - the trouble is that it does not always draw the boundary clearly. Readers, as a result, must judge for themselves whether "AI evaluates" denotes a function or whether "educators must verify" expresses an obligation.

This asymmetry is defensible on educational grounds. A system cannot be held to defend a degree standard, suffer a loss of professional standing, or correct a learner's miseducation. Problems arise, however, when humans carry accountability without sufficient control. Learners, for instance, may be told to verify GenAI-generated responses without access to sources, domain knowledge, or disclosure expectations. Educators, similarly, may be held accountable for assessing the authenticity of GenAI-generated responses without adequate institutional time, policy consistency, or control over the platforms used to procure GenAI.

### 5.2 Responsibility Recognized but Incompletely Relational

There is substantial evidence that the discipline recognizes responsibility; the concern is that relational completeness is lacking. Terms like "responsible AI", "integrity",

"governance", and "verification" each point to several possible objects: safe model design, lawful procurement, ethical instruction, fair student use of GenAI, transparency in research, and equitable access to GenAI. "Responsibly using GenAI", conversely, assigns duties to users while leaving provider and institutional choices outside the sentence entirely.

The obligation analysis clarifies this pattern. Humans and institutions predominate, yet roughly one-fifth of obligations were impersonal, and roughly half were expressed through policy, curriculum, or assessment as metonyms for an obligor. That phrasing can be convenient, but it obscures the chain linking obligation to action, which matters because governance failures often arise across levels. An undisclosed use by a learner can point to an unclear task; an unclear task can point to conflicting faculty directions; and conflicting faculty directions can point to a policy that states principles without assigning operational ownership.

Finally, the near absence of providers deepens this individualization. Providers - developers and vendors - shape training data, interfaces, privacy conditions, model updates, and access, yet they were the least visible actor group of all. The literature's human-AI binary can thus obscure organizations on both sides: commercial providers fade into "AI", while committees and administrators fade into "the institution". Ecological responsibility, as a result, narrows into a relationship between an abstraction called a "system" and an individual user.

**5.3 Variation by Representation**

The four representations develop this point further. Tool discourse contained the most AI-action predicates; calling a system a "tool" does not stop us from describing it in strongly agentive language, since "tool" descriptions are generally lists of capabilities. Interlocutor discourse produced the fewest AI-action predicates, yet it also included examples where dialogue facilitated reflective judgment. Proxy discourse, in contrast, placed the greatest relative weight on student evaluation and verification, since substitution requires demonstrating evidence of ownership. Infrastructure discourse, finally, used the largest proportion of nominalizations, since policies and governance are often represented as systems of abstract requirements.

These variations show that responsibility cannot be read off labels alone. A "partner" could remain tightly constrained by a learner's explicit judgment, while a "tool" could quietly carry out most of the epistemic work. A proxy could be acceptable if the point is to evaluate the proxy's output. Infrastructure, conversely, can either promote agency through contested decisions or eliminate it through automated enforcement. Analysis should therefore ask what operation is actually being carried out, who sets the standards, who can dispute the results, and who bears the consequences.

**5.4 Contributions to Research on Epistemic Agency**

The theoretical contribution here is a four-part conceptual distinction drawn from empirical observation: grammatical activation, functional agency, epistemic authority, and normative accountability. Each category relates to the others, but is not equivalent. Grammatical activation makes an actor identifiable. Functional agency refers to performance that produces consequences. Authority refers to a warranted capacity to judge and endorse. Accountability refers to an obligation to justify and to bear consequences.

This distinction connects the paper to broader work on learner-AI interaction without duplicating prior level-based research on how responsibility moves across an event sequence - that remains an open question, not one this study needs to resolve. Instead, the present findings address a different question: Does the existing literature offer adequate language to describe how responsibility moves and where it sits institutionally? The answer is mixed. Scholarship on verification, disclosure, and governance continues to grow, but much of it bundles these concerns together under broad quality- or nominal-process rubrics. So, although the educational literature has developed an extensive ethical vocabulary, it still lacks a clearly articulated, conventional language for identifying who assumes practical accountability.

## 6. Implications for Research, Pedagogy, and Governance

Claims in research reports should assign responsibility through an auditable relationship. At minimum, such claims should include the responsible actor, the object of knowledge or decision at stake, the actions to be performed, the applicable standard of evidence or professionalism, the audience to whom justification is owed, and the consequence - punitive or reparatory - for failing to meet the requirement.

Rather than write "verify AI outputs", a paper could say that students must verify their factual assertions and citations before submission, using specific primary or scholarly references. Educators would then set the criteria for acceptable evidence, and institutions would develop procedures for handling disputes over the validity of submitted work.

From a pedagogical standpoint, verification should not be treated as a last-minute obligation dropped on students once AI has produced content. It is instead a teachable skill, one that includes assessing sources, holding domain knowledge, understanding uncertainty, and exercising judgment. Designs built around doubt-based and fact-checking approaches show how students' use of AI-generated content can become an object of reasoning rather than a shortcut around it (Ismail et al., 2026). Studies of students' revision practices further show that acceptance, modification, and rejection are distinct forms of agency (Jiang et al., 2026). Assessments should reflect these differences by drawing on evidence from the revision process itself, students' commentary and defense of revised work, and comparisons between original and revised work. Assessments should not, in short, ask students to disclose AI use and stop there.

Policymakers and institutional administrators should treat responsibility as a function of control when drafting AI-accountability policy. Providers, for instance, can be held accountable for the documented characteristics of their systems and for contractual assurances about system quality; institutions can be held accountable for the procurement process, the consistency of institution-wide policy, the resources and support given to educators, and the decisions institutions themselves make; educators can be held accountable for task design and feedback within the authority the institution grants them; and students can be held accountable for evidence-based judgments once given sufficient training and education to make them (Ucan, 2026). Detection scores should prompt investigation rather than serve as automatic determinations of acceptability (Ucan, 2026). Accessibility deserves careful attention, too: mediated expression can protect authors' intellectual property rights rather than diminish them (Morgan, 2026).

Finally, researchers should write about AI clearly and concisely, favouring the active voice. A phrase like "the user utilizes AI to…" does little beyond obscuring what the AI system actually did. A stronger approach adds a normative clause that states what the system does, what it cannot guarantee or justify, and who decides whether its output is acceptable.

## 7. Limitations and Future Research

The dataset covers titles and abstracts only. These summarize an article's main points but carry far less information about method, about who did what in joint authorship, or about how different parts of an article relate to one another - which section supplies the evidence, for instance. Title-weighted filtering, therefore, favours work that explicitly names who or what is responsible for an action. All available records from Scopus and Web of Science were imported using every available filter; because IEEE does not allow filtered exports without membership, only IEEE journal articles were included. Springer, which allows relevance-ranked search for 2024–2026, was searched across the entire ranked set using that method. Forty-eight articles identified in the Springer export lacked corresponding abstracts and were excluded. The resulting corpus is large, but it is still probably not exhaustive.

The corpus skews heavily toward 2026, a year that is both partially complete and still rapidly evolving. Beyond the general problem of using an incomplete year in trend analysis, a second problem involves online-first and issue/year date fields: when an article appears online before its official issue date, the recorded "date" can lag the actual publication date. Similarly, some issues carry a date tied to when the final version was published, while others carry the date of the new year in which they were assigned. The rules used to assign actor-predicate relationships are approximations and do not capture longer-range dependencies such as pronoun reference or negation. Using passive-voice cues to detect obligation relationships can surface oblique constructions that would otherwise go undetected. Operational-representation coding simplifies documents that combine multiple content types, such as data, tables, and figures, and publication-type coding adds a further layer of abstraction.

Several additional studies should precede firm confidence in these conclusions. First, future work should build a legally accessible subset of the corpus consisting of full-text articles. Second, it should verify the accuracy of the actor-predicate rules through two approaches: dependency parsing, to check whether the rule-based approach identifies the same relationships a parser would, and double coding, to check whether the findings hold when another researcher codes the same material. Third, it should examine whether responsibility language in abstracts repeats elsewhere in the article - that is, whether abstract language represents the language of the full article - which would help determine whether responsibility language becomes more frequent once the discussion moves past the abstract.

## 8. Conclusion

A high-functioning system carries considerable functional agency in its roles as generator, supporter, and transformer. However, the higher education community does not treat GenAI as normatively equivalent to either human or institutional actors. Students, accordingly, are more likely to know, judge, and verify than to act with other tools; educators and institutions bear most of the obligation; providers remain on the periphery; and much of what occurs occurs through passive or nominal language. The community recognizes "responsibility", in other words, but at times it blurs the relationship that turns recognition into action.

None of this removes agency from our definition of GenAI. It does mean that scholarship should keep the difference between performing an action, holding authority over an action, and being accountable for an action clearly in view. Once the discipline articulates these distinctions, it becomes easier to say which party acted, which party warranted the knowledge claim, which party could dispute the decision, and which party will answer for the consequences. For institutions using GenAI to produce educational work, this distinction may ultimately decide whether the resulting products - made by humans and systems together - remain both academically intelligible and defensible.

## References

Alotaibi, A. E., Aseery, A., Alfayez, A. A., Bashatah, L., & Aseri, S. (2026). Governance of generative AI in higher education: A qualitative analysis of universities' policies and global insights. *British Educational Research Journal*. https://doi.org/10.1002/berj.70256

Atamhenwan, L., Kutty, S., & Li, L. D. (2026). Defining responsible AI: Contextual insights powered by LLMs. In *Communications in Computer and Information Science* (pp. 224–237). Springer. https://doi.org/10.1007/978-981-95-6786-7_15

Baker, P., Gabrielatos, C., Khosravinik, M., Krzyżanowski, M., McEnery, T., & Wodak, R. (2008). A useful methodological synergy? Combining critical discourse analysis and corpus linguistics to examine discourses of refugees and asylum seekers in the UK press. *Discourse & Society, 19*(3), 273–306. https://doi.org/10.1177/0957926508088962

Bozkurt, A., Crompton, H., & Kurban, C. F. (2026). The devil is in the det[ai]ls: AI agents, ghost students, and the crisis of verified presence in an agentic AI world. *Open Praxis, 18*(1), 1–12. https://doi.org/10.55982/openpraxis.18.1.1145

Coeckelbergh, M. (2020). Artificial intelligence, responsibility attribution, and a relational justification of explainability. *Science and Engineering Ethics, 26*, 2051–2068. https://doi.org/10.1007/s11948-019-00146-8

Cong-Lem, N. (2025). Exploring human–AI distributed critical thinking (HADCT): Pilot validation of a critical thinking scale with Vietnamese EFL learners. *TEFLIN Journal, 36*(2), 279–302. https://doi.org/10.15639/teflinjournal.v36i2/279-302

Crawford, J., Cowling, M., Ashton-Hay, S., Kelder, J.-A., Middleton, R., & Wilson, G. S. (2023). Artificial intelligence and authorship editor policy: ChatGPT, Bard, Bing AI, and beyond. *Journal of University Teaching and Learning Practice, 20*(5). https://doi.org/10.53761/1.20.5.01

Dai, Y., Liu, S., Zhou, S., Lai, S., Liu, A., & Lim, C. P. (2026). Redefining and measuring student agency in AI-assisted learning: Development and validation of the agentic engagement with AI (AE-AI) scale. *Computers & Education, 253*, 105687. https://doi.org/10.1016/j.compedu.2026.105687

De Maio, C. (2024). Institutional responses to ChatGPT: Analysing the academic integrity policies of four public and private institutions of higher education in Australia. *Journal of Academic Language and Learning, 18*(1), T1–T8.

Essien, A., Zhou, X., Kremantzis, M., & Teng, D. (2026). The agency gap: Perceived human–AI agency, reflection and generative AI learning across UK- and China-based higher education contexts. *Studies in Higher Education*, 1–22. https://doi.org/10.1080/03075079.2026.2686986

Fairclough, N. (1992). *Discourse and social change*. Polity Press.

Floridi, L., & Sanders, J. W. (2004). On the morality of artificial agents. *Minds and Machines, 14*, 349–379. https://doi.org/10.1023/B:MIND.0000035461.63578.9d

Halliday, M. A. K., & Matthiessen, C. M. I. M. (2014). *Halliday's introduction to functional grammar* (4th ed.). Routledge. https://doi.org/10.4324/9780203431269

Ismail, A. F., Neyef Al-Saliti, R. A. M., Al-Muoaeweed, O., & Mohamed Hamid, A. A. (2026). The effectiveness of a doubt-based learning program in enhancing critical thinking and self-efficacy in verifying AI-generated knowledge among university students. *Journal of Educational and Social Research, 16*(3), 327. https://doi.org/10.36941/jesr-2026-0335

Jiang, Y., Wu, Q., Yang, Y., Jian, C., & Zhao, J. (2026). Learner agency in revising GenAI-generated statements of purpose. *British Journal of Educational Technology, 57*(4), 965–983. https://doi.org/10.1111/bjet.70041

Karam, J. G. (2026). Between panic and hype: Disclose and defend pedagogy for AI writing in political science classrooms in the Global South. *PS: Political Science & Politics*, 1–8. https://doi.org/10.1017/S1049096526102261

Lindebaum, D., Nolan, E., Ashraf, M., Islam, G., & Ramirez, M. F. (2026). The transformation of epistemic agency and governance in higher education through large language models: Toward a future of organized immaturity. *Organization Studies*. https://doi.org/10.1177/01708406251392002

Liu, Y., Jou, M., & Hao, Y. (2026). From cognitive offloading to learning accountability erosion: How generative AI reshapes students' epistemic trust and educational engagement. *Interactive Learning Environments*, 1–17. https://doi.org/10.1080/10494820.2026.2680276

Madsen, D. Ø., & Silva, E. S. (2026). Teaching citation in the age of generative AI: Rethinking research literacy, academic integrity, and epistemic responsibility. *The Journal of Academic Librarianship, 52*(4), 103267. https://doi.org/10.1016/j.acalib.2026.103267

Moorhouse, B. L., Wan, Y., Wu, C., Wu, M., & Ho, T. Y. (2025). Generative AI tools and empowerment in L2 academic writing. *System, 133*, 103779. https://doi.org/10.1016/j.system.2025.103779

Morgan, D. (2026). Reasonable adjustments for neurodivergent students in higher education: Generative AI, accessibility, and mediated authorship. *Disability & Society, 41*(7), 1836–1856. https://doi.org/10.1080/09687599.2026.2667528

Page, M. J., McKenzie, J. E., Bossuyt, P. M., Boutron, I., Hoffmann, T. C., Mulrow, C. D., Shamseer, L., Tetzlaff, J. M., Akl, E. A., Brennan, S. E., Chou, R., Glanville, J., Grimshaw, J. M., Hróbjartsson, A., Lalu, M. M., Li, T., Loder, E. W., Mayo-Wilson, E., McDonald, S., … Moher, D. (2021). The PRISMA 2020 statement: An updated guideline for reporting systematic reviews. *BMJ, 372*, n71. https://doi.org/10.1136/bmj.n71

Perkins, M. (2023). Academic integrity considerations of AI large language models in the post-pandemic era: ChatGPT and beyond. *Journal of University Teaching and Learning Practice, 20*(2). https://doi.org/10.53761/1.20.02.07

Perkins, M., & Roe, J. (2024). Decoding academic integrity policies: A corpus linguistics investigation of AI and other technological threats. *Higher Education Policy, 37*(3), 633–653. https://doi.org/10.1057/s41307-023-00323-2

Qu, Y., & Wang, J. (2026). To disclose or not to disclose: Peer influence and psychological factors in students' use of generative artificial intelligence. *British Journal of Educational Psychology*. https://doi.org/10.1111/bjep.70086

Qutieshat, A., Annamma, L. M., Singh, G., Arzmi, M. H., Wan Ahmad Kamil, W. N., Khasawneh, L., Varma, S. R., Carneiro Leão, J., George, B. T., & Alrashdan, M. S. (2026). Large language model use in dental education: A cross-sectional multi-country study. *Medical Education Online, 31*(1), 2707729. https://doi.org/10.1080/10872981.2026.2707729

Rauschenbach, I. (2026). Using AI as a teaching partner: Enhancing critical thinking and digital literacy in microbiology. *Journal of Microbiology & Biology Education*, e00326-25. https://doi.org/10.1128/jmbe.00326-25

Si, S., Qi, Y., Xu, J., & Qi, X. (2026). When thinking is outsourced: Cognitive offloading and the heterogeneity of critical thinking among Chinese university students using generative artificial intelligence. *Journal of Intelligence, 14*(7), 116. https://doi.org/10.3390/jintelligence14070116

Sporrong, E., McGrath, C., Viberg, O., & Cerratto Pargman, T. (2025). What is the problem with generative artificial intelligence in higher education? A critical analysis of educator responsibility in the Swedish policy landscape. *Learning, Media and Technology*, 1–20. https://doi.org/10.1080/17439884.2025.2607687

Sullivan, M., Kelly, A., & McLaughlan, P. (2023). ChatGPT in higher education: Considerations for academic integrity and student learning. *Journal of Applied Learning & Teaching, 6*(1). https://doi.org/10.37074/jalt.2023.6.1.17

Tan, X. (2026). "I have not felt AI writing is taking away my agency as a person": A longitudinal study of ESL students' use of GenAI in academic literacies development. *System, 141*, 104099. https://doi.org/10.1016/j.system.2026.104099

Tyndall, E., Gayheart, C., Some, A., Genz, J., Wagner, T., & Langhals, B. (2025). Impact of retrieval augmented generation and large language model complexity on undergraduate exams created and taken by AI agents. *Data & Policy, 7*, e57. https://doi.org/10.1017/dap.2025.10024

Ucan, S. (2026). Beyond the detector trap: A human-centred evidentiary framework for fair academic integrity decisions in the generative AI era. *Assessment & Evaluation in Higher Education*, 1–16. https://doi.org/10.1080/02602938.2026.2711422

Urban, M., Brom, C., Lukavský, J., Děchtěrenko, F., Hein, V., Svacha, F., Kmoníčková, P., & Urban, K. (2025). "ChatGPT can make mistakes. Check important info." Epistemic beliefs and metacognitive accuracy in students' integration of ChatGPT content into academic writing. *British Journal of Educational Technology, 56*(5), 1897–1918. https://doi.org/10.1111/bjet.13591

van Leeuwen, T. (2008). *Discourse and practice: New tools for critical discourse analysis*. Oxford University Press. https://doi.org/10.1093/acprof:oso/9780195323306.001.0001

Willems, T., Khan, S., Huang, Q., Camburn, B., Sockalingam, N., & Poon, K. W. (2025). To use or to refuse? Re-centering student agency with generative AI in engineering design education. In *2025 IEEE International Conference on Teaching, Assessment, and Learning for Engineering* (pp. 1–8). IEEE. https://doi.org/10.1109/TALE66047.2025.11346653

Wu, E. (2026a). From integrity to identity: Course-level generative AI governance and scholarly subject formation in graduate education. *Frontiers in Education, 11*, 1827251. https://doi.org/10.3389/feduc.2026.1827251

Wu, E. (2026b). From production to verification: Generative AI, doctoral formation, and the leadership of digital education. *Frontiers in Education, 11*, 1890769. https://doi.org/10.3389/feduc.2026.1890769